\documentclass[a4paper,fleqn,usenatbib]{mnras}

\usepackage{newtxtext,newtxmath}

\usepackage[T1]{fontenc}
\usepackage{ae,aecompl}

\usepackage{graphicx}	
\usepackage{amsmath}	
\usepackage{amssymb}	
\usepackage{bm}
\usepackage{dsfont}
\usepackage{braket}
\usepackage{cprotect}

\title[Anomaly detection for X-ray spectroscopy]{Neural network-based anomaly detection for high-resolution X-ray spectroscopy}

\author[Y. Ichinohe and S. Yamada]{
Y. Ichinohe,$^{1,2}$\thanks{E-mail: ichinohe@rikkyo.ac.jp}
and S. Yamada,$^{3}$
\\
$^{1}$Department of Physics, Rikkyo University, 3-34-1 Nishi-Ikebukuro, Toshima, Tokyo 171-8501, Japan\\
$^{2}$Bluish AI Laboratory, 7-11-3 Ginza, Chuo, Tokyo 104-0061, Japan\\
$^{3}$Department of Physics, Tokyo Metropolitan University, 1-1 Minami-Osawa, Hachioji, Tokyo 192-0397, Japan\\
}

\date{\today}

\pubyear{2019}

\begin{document}
\label{firstpage}
\pagerange{\pageref{firstpage}--\pageref{lastpage}}
\maketitle

\begin{abstract}
We propose an anomaly detection technique for high-resolution X-ray spectroscopy. The method is based on the neural network architecture variational autoencoder, and requires only {\it normal} samples for training. We implement the network using Python taking account of the effect of Poisson statistics carefully, and deonstrate the concept with simulated high-resolution X-ray spectral datasets of one-temperature, two-temperature and non-equilibrium plasma. Our proposed technique would assist scientists in finding important information that would otherwise be missed due to the unmanageable amount of data taken with future X-ray observatories.
\end{abstract}




\section{Introduction}
Most of the baryons in the Universe are in the form of diffuse plasma \citep{shull12}. The ionised gas produces bremsstrahlung continua and fluorescent emission lines, and absorbs the X-rays corresponding to atomic transitions. The spectral features contain fruitful information on various plasma properties; e.g., thermodynamic properties such as temperature and density, chemical properties such as metal abundances, plasma properties such as ionisation states, and geometrical properties such as redshift and line-of-sight velocity dispersion. As these features, in particular the ones originating from plasma that is heated up to 10$^{7-8}$~K by various energetic phenomena appear mainly in the X-ray wavelengths, X-ray spectroscopy is one of the most essential tools regarding the cosmic baryon study.

For years, X-ray CCDs and grating spectrometers have been used for X-ray spectroscopy. As CCD has imaging capability, one can perform imaging spectroscopy with moderate energy resolutions of $\Delta E\sim100~\mathrm{eV}$. On the other hand, grating spectrometers have high energy resolutions of $\Delta E/E\sim1000$, but are not suitable for diffuse objects because the spatial extent of the targets degrades the energy resolution. Therefore, high energy resolution spectroscopy of diffuse astrophysical objects has been difficult.

X-ray spectroscopy is entering a new era with the advent of a cryogenic X-ray microcalorimeter. X-ray microcalorimeters measure the energy of each photon as the temperature change of the absorber material. They are operated under low temperatures ($\sim$100~mK) in order to suppress thermal noises and heat capacities, and the resulting resolution power reaches $\Delta E/E\sim1000$. Because microcalorimeters are non-dispersive detectors, spatially resolved spectroscopy of diffuse targets is also possible using pixelised calorimeter arrays and X-ray focusing mirrors.

The first attempts with an X-ray microcalorimeter started with the XQC sounding rocket experiment launched in 1995, resulting in a successful five-minute observation \citep{porter05}. By learning essential lessons from the subsequent space missions, the first X-ray microcalorimeter in orbit was realised by the Soft X-ray Spectrometer \citep[SXS;][]{kelley16} onboard the {\it Hitomi} satellite \citep{takahashi18}. Before the sudden loss of the attitude control, {\it Hitomi} observed several targets including the Perseus cluster of galaxies. SXS successfully detected its plasma emission and demonstrated the superb capability of X-ray microcalorimeters for astrophysical plasma observations \citep[e.g.][]{hitomi16,hitomidm,hitomiz,hitomiv,hitomirs,hitomit,hitomiatomic}.

The information uncovered by {\it Hitomi} is only the tip of the iceberg. {\it Hitomi} pioneered the operation of a semiconductor microcalorimeter in space, and its recovery mission {\it XRISM} \citep{tashiro18} will resume {\it Hitomi}'s science advancement and mature its technologies. The semiconductor microcalorimeter will lead the application of the next generation microcalorimeters, superconducting transition-edge sensors (TES). TESs exploit the multiplex readout technology using superconducting quantum interference devices (SQUID), which match the low impedance of the detectors and realise better energy resolution than semiconductor-based microcalorimeters. TES is a key technology for spatially resolved spectroscopy with high spectral resolution that will allow significant advances in high-energy astrophysics. {\it Athena} \citep{barcons15} will be the first mission which aims at operating TESs in space. The expected number of pixels is of order 10$^3$. The future X-ray missions such as {\it Lynx} \citep{ozel18} and {\it SuperDIOS} \citep{ohashi18} envision $>10^3$ pixels.

Analysing the huge amount of high-resolution X-ray spectra brought by future satellites will be almost impossible considering the available resources (time, human and computational). Countermeasures such as parallel processing using central processing unit (CPU) or graphics processing unit (GPU) might alleviate the situation, but most of the expected issues can not be easily solved by just dividing the tasks. Individual tasks need {\it intelligence} which can adjust strategies according to the situation; e.g., optimization, regression or classification. Therefore, the need for {\it intelligent} automated data analysis frameworks is becoming even more critical.

In such a situation, deep learning is gaining popularity even in the field of astronomy \citep[e.g.][]{hezaveh17,daniel18,schaefer18,leung19}. Similar to other machine learning techniques, it trains a computer to make {\it intelligent} judgements and recognize the pattern embedded in the data. For example in \citet{schaefer18}, a deep learning architecture convolutional neural network (CNN) is proposed as the method to characterize a large amount of galaxy images, and tell us whether each individual galaxy is affected by gravitational lensing or not.

Recently, \citet{ichinohe18} demonstrated that the technique of deep learning can actually be applied to the parameter estimation problem of X-ray spectroscopy, under the assumption that the model is known and simple (i.e. single-temperature plasma in collisional ionisation equilibrium). However, the assumption might be actually too simple considering the diversity of plasma objects. For example, non-equilibrium plasma is often observed in supernova remnants \citep[see e.g.][for a review]{vink12}, and the spectrum may consist of multiple plasma components of different temperatures due to, e.g., the projection effect \citep{ettori02}. Importantly, it is often the case that the spectra which cannot be modelled under simple assumptions are scientifically more important because this implies the existence of factors beyond those included in the simple assumption.

In this paper, we propose a new, automated method that can detect such {\it anomalies} in X-ray spectra. In Section~\ref{sec:vae}, we briefly describe the concept of our new method in which a neural network architecture, variational autoencoder \citep[VAE;][]{kingma13} is applied. The actual network implementation and the demonstration using simulated datasets are shown in Section~\ref{sec:demo}. We discuss the results with prospects in Section~\ref{sec:discussion}.

\section{Anomaly detection using VAE}\label{sec:vae}
Here we briefly describe the concept of the anomaly detection using the neural network architecture VAE. For elaborate explanations of neural networks and deep learning, see e.g. \citet{lecun15} and \citet{goodfellow16}. For a recent review regarding anomaly detection using deep learning, see \citet{chalapathy19}.

\subsection{Variational autoencoder}
Our method is based on a deep neural network architecture, variational autoencoder \citep[VAE;][]{kingma13}, which is a variant of another network architecture, autoencoder \citep[AE;][]{hinton06}. Basically, an AE consists of an encoder and a decoder, and takes a feedforward neural network architecture that consists of several fully-connected layers; as the input vector of dimension $d$ is fed forward in the encoder of the network, the dimension of a layer (i.e. the number of nodes or {\it neuron}s in each layer) is gradually reduced to a number smaller than $d$ ($d_\mathrm{latent}$). This $d_\mathrm{latent}$-dimensional vector is referred to as the latent expression. The dimension of the latent expression is increased back to the original dimension $d$ through the decoder of the network. The whole AE network (, which is a serial combination of the encoder and the decoder) is simultaneously trained so that the output reproduces the input. Once the network is successfully trained, one can obtain the $d_\mathrm{latent}$-dimensional latent expression for an arbitrary $d$-dimensional input vector. Since $d_\mathrm{latent}<d$, the latent expression can be regarded as a compressed representation of the input vector in a $d_\mathrm{latent}$-dimensional space.

From the point of view of the network architecture, VAE is modified from AE in the folloing points; (1) unlike the normal AE architecture where all the layers are connected deterministically, the latent expression of VAE is computed in a probabilistic sampling layer. The output of this layer is generated via random sampling from the multidimensional Gaussian distribution whose parameters are determined by the inputs to this layer; (2) the loss function includes not only the reconstruction error but also a regularisation term that arranges the distribution of the latent expressions.

From the point of view of Bayesian statistics, VAE is a tool to perform the variational Bayesian method to infer $p_{\bm{\theta}}({\bf z})p_{\bm{\theta}}({\bf x}|{\bf z})$, the generative model of ${\bf x}$ with the latent variables ${\bf z}$ that maximises the marginal likelihood $p_{\bm{\theta}}({\bf x})$ for a given training dataset of ${\bf x}$, where $\bm{\theta}$ are the generative model parameters. Usually the marginal likelihood $p_{\bm{\theta}}({\bf x}) = p_{\bm{\theta}}({\bf z})p_{\bm{\theta}}({\bf x}|{\bf z})/p_{\bm{\theta}}({\bf z}|{\bf x})$ is intractable, and thus in order to estimate the true $p_{\bm{\theta}}({\bf x})$, some approximations are applied to the exact form.

Let $q_\phi({\bf z}|{\bf x})$ be the approximation to the intractable posterior $p_{\bm{\theta}}({\bf z}|{\bf x})$, the variational lower bound $\mathcal{L}(\bm{\theta},\bm{\phi};{\bf x}^{(i)})$ is expressed as:
\begin{equation}
\mathcal{L}(\bm{\theta},\bm{\phi};{\bf x}^{(i)})=-D_\mathrm{KL}(q_\phi({\bf z}|{\bf x}^{(i)}) \parallel p_{\bm{\theta}}({\bf z})) + \mathds{E}_{q_\phi({\bf z}|{\bf x}^{(i)})}\left[\log p_{\bm{\theta}}({\bf x}^{(i)}|{\bf z})\right],\label{eq:vlb}
\end{equation}
where the first term is the Kullback-Leibler divergence of $q_\phi({\bf z}|{\bf x})$ from $p_{\bm{\theta}}({\bf z})$, and the second term is the expected value of the log-likelihood of the $i$-th datapoint ${\bf x}^{(i)}$ over the probability distribution $q_\phi({\bf z}|{\bf x}^{(i)})$. As $\log p_{\bm{\theta}}({\bf x})\geq \mathcal{L}(\bm{\theta},\bm{\phi};{\bf x})$ always holds, maximising $\mathcal{L}(\bm{\theta},\bm{\phi};{\bf x})$ is a good alternative of directly maximising $p_{\bm{\theta}}({\bf x})$.

$q_\phi({\bf z}|{\bf x})$ and $p_{\bm{\theta}}({\bf x}|{\bf z})$ can be realised simultaneously using an autoencoder-type neural network as folloing. First, one can assume the functional form of $q_\phi({\bf z}|{\bf x})$, i.e. the approximation to the true posterior, as a multivariate Gaussian $\mathcal{N}({\bf z};{\boldsymbol \mu},{\boldsymbol \sigma^2}{\bf I})$, where ${\boldsymbol \mu}$ and ${\boldsymbol \sigma^2}{\bf I}$ are the mean and the (diagonal) covariance matrix of the multivarite Gaussian, which are computed through the encoder of the neural network. Similarly, the likelihood $p_{\bm{\theta}}({\bf x}|{\bf z})$ is modelled using a probability distribution whose parameters are determined using the decoder of the neural network.

In this implementation, the second term of Eq.~\ref{eq:vlb} corresponds to the reconstruction error of the autoencoder network, while the first term acts as a reguraliser of the distribution of latent expression ${\bf z}$ in the $d_\mathrm{latent}$-dimensional space. Training the VAE network is therefore the same as learning a generative model, which approximates the true generative model distribution from which the training data are generated.

\subsection{Anomaly detection}
Suppose a VAE network has been successfully trained using a given dataset. The decoder of the network has the ability to generate the mock data instance corresponding to a given latent expression from the generative model of the training dataset.

When inputs are fed into the network, the corresponding outputs are twofold; (1) if the input is generated from the same generative model as that of the training dataset, the VAE output is similar to the input because the network is trained so. However, (2) if the input is generated from a different generative model, the VAE output is likely different from the input, because the input is generated from a generative model that is not the one that the decoder of the network approximates. Therefore, by comparing the VAE output to the input, one can detect the inputs whose generative model is different from the training generative model. In other words, one can detect {\it anomalies} in the inputs, assuming the inputs that are generated from the training generative model to be {\it normal}.

\section{Demonstration}\label{sec:demo}
We constructed an actual VAE network to illustrate the concept shown in the previous section using X-ray high-resolution spectroscopic datasets for demonstration.

\subsection{Implementation}\label{sec:impl}
\begin{figure*}
  \centering
  \includegraphics[width=5.5in]{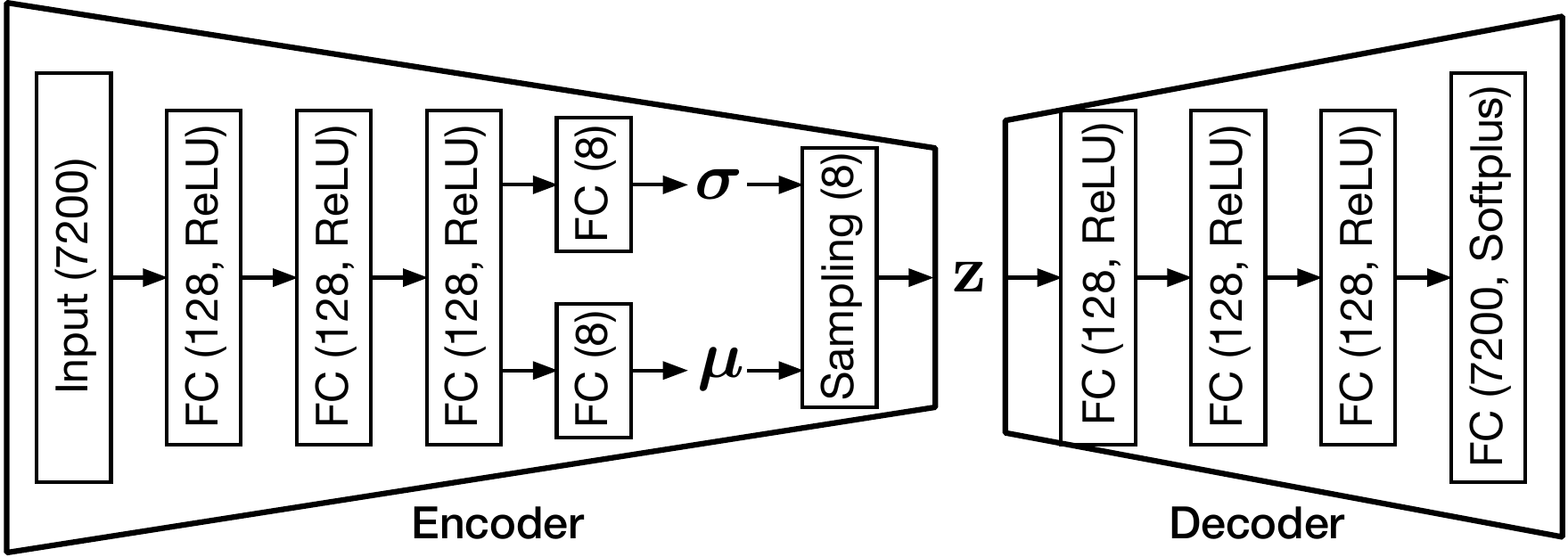}
 \caption[]{Conceptual diagram of the variational autoencoder network used in this work.}
 \label{img:network}
\end{figure*}
Fig.~\ref{img:network} shows the conceptual diagram of the network adopted in this work. Several fully-connected (FC) layers and one sampling layer are present in the encoder network. A fully-connected layer transforms the input $d_{\bf x}$-dimensional vector ${\bf x}$ to the output $d_{\bf y}$-dimensional vector ${\bf y}$ using ${\bf y} = f({\mathrm W}{\bf x} + {\bf b})$, where ${\mathrm W}$ and ${\bf b}$ are a $d_{\bf y}\times d_{\bf x}$ matrix and  a $d_{\bf y}$-dimensional vector, respectively, whose elements are optimised in the training, and $f(\cdot)$ is an element-wise nonlinear transformation referred to as an activation function.

In our case, the input to the network is an X-ray spectrum extracted in the energy range of 1.8--9.0~keV with the energy bin size of 1~eV (i.e. a 7200-dimensional vector). The output dimension of the first three fully-connected layers are 128 (i.e. $d_{\bf y}=128$) and rectified linear unit \citep[ReLU;][]{nair10} is used for the activation function of these layers. The output of the third layer is connected to two independent fully-connected layers whose output dimension $d_{\bf y}=8$ and whose outputs are not activated (i.e. $f(x)=x$). These two 8-dimensional outputs are fed into the sampling layer.

The sampling layer randomly generates an 8-dimensional latent vector ${\bf z}$ from an 8-dimensional multivariate Gaussian distribution; ${\bf z}\sim \mathcal{N}({\boldsymbol \mu},{\boldsymbol \sigma^2}{\bf I})$, where $\boldsymbol \mu$ or $\boldsymbol \sigma$ is represented by each of the two input 8-dimensional vectors.

The decoder part of the network accepts the latent vector ${\bf z}$ as the input and comprises four fully-connected layers. The output dimension of the first three layers are also 128 and ReLU activation is used. The output dimension of the last layer is 7200, which is the same as the original input dimension.

The activation function of the last layer should be carefully chosen so as to be consistent with the physical process to obtain the dataset. In our case, the dataset consists of X-ray spectra, and the value of each bin in an X-ray spectrum represents the number of photons, which is a Poissonian random variable. Therefore, the output of the last layer should be consistent with being the expectation value of a Poission distribution, which means that it should be a positive real number. We therefore used a softplus function $f(x)=\log (1+e^x)$ since its range is $f(x)>0$ (c.f. $f(x)\geq 0$ for ReLU, $0<f(x)<1$ for a sigmoid function and $-1<f(x)<1$ for a $\tanh$ function).

Assuming that the prior distribution is a standard Gaussian $p_{\bm{\theta}}({\bf z}) = \mathcal{N}({\mathrm 0},{\bf I})$, the first term of the loss function (Eq.~\ref{eq:vlb}) can be explicitly expressed as
\begin{equation}
-D_\mathrm{KL}(q_\phi({\bf z}|{\bf x}^{(i)}) \parallel p_{\bm{\theta}}({\bf z}))=\dfrac{1}{2}\sum_{j=1}^{8}(1+\log((\sigma_j^{(i)})^2)-(\mu_{j}^{(i)})^2-(\sigma_{j}^{(i)})^2),
\end{equation}
where $\sigma_{j}^{(i)}$ or $\mu_{j}^{(i)}$ is the $j$th element of ${\boldsymbol \sigma}$ or ${\boldsymbol \mu}$ calculated for the $i$th input of the dataset.

The second term is the poisson reconstruction error in our case;
\begin{equation}
\begin{split}
\mathds{E}_{q_\phi({\bf z}|{\bf x}^{(i)})}\left[\log p_{\bm{\theta}}({\bf x}^{(i)}|{\bf z})\right]\simeq
\sum_{j=1}^{7200} \log\dfrac{e^{-{\bf y}_{j}^{(i)}} {{\bf y}_{j}^{(i)}}^{{\bf x}_{j}^{(i)}}}{{{\bf x}_{j}^{(i)}}!}\\
\sim \sum_{j=1}^{7200} \left( -{\bf y}_{j}^{(i)} + {\bf x}_{j}^{(i)}\log{\bf y}_{j}^{(i)} - {\bf x}_{j}^{(i)}\log{\bf x}_{j}^{(i)} + {\bf x}_{j}^{(i)} - \dfrac{\log{\bf x}_{j}^{(i)}}{2} \right),\label{eq:loss}
\end{split}
\end{equation}
where ${\bf y}_{j}^{(i)}$ is the $j$th value of the decoder output ${\bf y}^{(i)}$ corresponding to the $i$th input ${\bf x}^{(i)}$, and Stirling's approximation is used.

We implemented the network using the Python neural network library Keras \citep{chollet15} with the TensorFlow backend \citep{abadi15}.\footnote{Sample codes are available at: \url{https://github.com/yutoichinohe/sample_codes}.}

Bigger networks are expected to have more expressive power while training them is computationally more expensive. We thus tried other network architectures to explore the optimal complexity of the network. In addition to the network described above in which three fully-connected layers with $d_{\bf y}=128$ are implemented both in the encoder and the decoder (128$\times$3 network), we also tried 64$\times$2 and 256$\times$5 networks. We found that when we use the 64$\times$2 network, the attained value of the loss function is worse than that of 128$\times$3 network. On the other hand, when we use the 256$\times$5 network, the attained value is alsmost same as that of 128$\times$3 network. Therefore, our network architecture appears to be nearly optimal for the current training dataset described in the next section.

\subsection{Dataset}
We prepared the training dataset using X-ray spectral simulation following the same procedure described in \citet{ichinohe18}. We used the \verb+fakeit+ command in {\small XSPEC} (version 12.10.0c) and {\it Hitomi} response files to generate simulated spectra. For simplicity, we used the response files for point sources. The training dataset consists of fake X-ray spectra whose model is represented by a single-temperature thermal plasma in collisional ionization equilibrium attenuated by the Galactic absorption (\verb+TBabs*apec+). Only temperature ($kT=1.0-10.0$~keV), Fe abundance ($Z=0.1-1.5$~solar), redshift ($z=0.0-0.1$), and normalization ($N=0.01-1.0$) were variable in the simulation, and the hydrogen column density was fixed to 1.38$\times 10^{21}~\mathrm{cm}^{-2}$ \citep[coressponding to the Perseus value,][]{kalberla05}. 10000 spectra were generated and used as the training dataset.

Besides the one-temperature (1T) training dataset, two {\it anomaly} datasets were also prepared for the use in the demonstration (1000 spectra each); (i) a two-temperature (2T) plasma dataset, in which the spectra are modelled using \verb|TBabs*(apec+apec)|, with the temeprature ($kT_{1,2}=1.0-10.0$~keV), Fe abundance ($Z_{1,2}=0.1-1.5$~solar), redshift ($z_{1,2}=0.0-0.1$) and normalization ($N_{1,2}=0.01-1.0$) of each of the two plasma components being variable parameters (8 free parameters in total); (ii) a non-equilibrium (NEI) plasma dataset, in which the model \verb+TBabs*nei+ is used with 5 variable parameters: temperature ($kT=1.0-10.0$~keV), Fe abundance ($Z=0.1-1.5$~solar), ionization timescale ($n_{\mathrm e}t=10^{9}-10^{11}$~s~cm$^{-3}$), redshift ($z=0.0-0.1$), and normalization ($N=0.01-1.0$).

\subsection{Training}
We trained the network using the 1T training dataset. 80\% of the training dataset was actually used for the training and 20\% for testing. The {\it Adam} algorithm \citep[][]{kingma14} was used for the network optimization. The batch size was 32, and the training was repeated for 10000 epochs.

\begin{figure}
  \centering
  \includegraphics[width=3.3in]{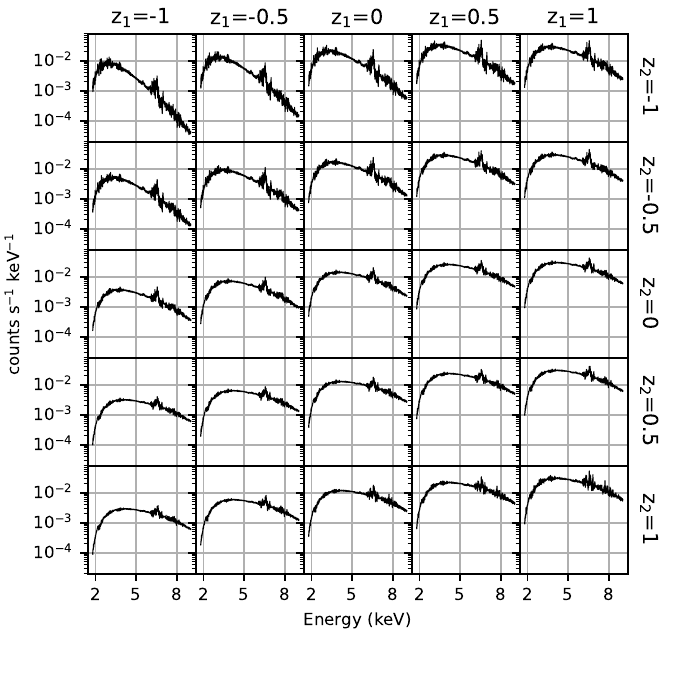}
 \caption[]{The learned data manifold visualized for a two-dimensional slice in the latent space. Each spectrum represents an 8-dimensional vector in the latent space. For example, the spectra shown in the top leftmost panel and the bottom rightmost panel are obtained by respectively feeding the latent expressions ${\bf z}=(-1,-1,0,0,0,0,0,0)$ and ${\bf z}=(1,1,0,0,0,0,0,0)$ to the decoder of the VAE network.}
\label{img:mani}
\end{figure}
Figure~\ref{img:mani} shows the learned data manifold visualized for a two-dimensional slice in the latent space. Each spectrum represents an 8-dimensional latent expression. The first component of the latent vector varies as -1, -0.5, 0, 0.5, and 1 from left to right, while the second component does likewise from top to bottom, with all the other elements fixed to zero. For example, the spectra shown in the top leftmost panel and the bottom rightmost panel are obtained by respectively feeding the latent expressions ${\bf z}=(-1,-1,0,0,0,0,0,0)$ and ${\bf z}=(1,1,0,0,0,0,0,0)$ to the decoder of the VAE network.

The normalizations of the spectra increase from left to right, while their slopes decrease (become harder) from top to bottom. This indicates that the first and second components roughly represent the normalization and temperature of the plasma model, respectively. Also importantly, the latent expressions are distributed smoothly in the latent space. This enables the decoder to interpolate the data instances that are not in the training dataset, and thus is essential for the decoder of the network to be a generative model, which is difficult to realise using ordinary AEs.

\subsection{Results}\label{sec:results}
\begin{figure*}
  \centering
  \includegraphics[width=7.0in]{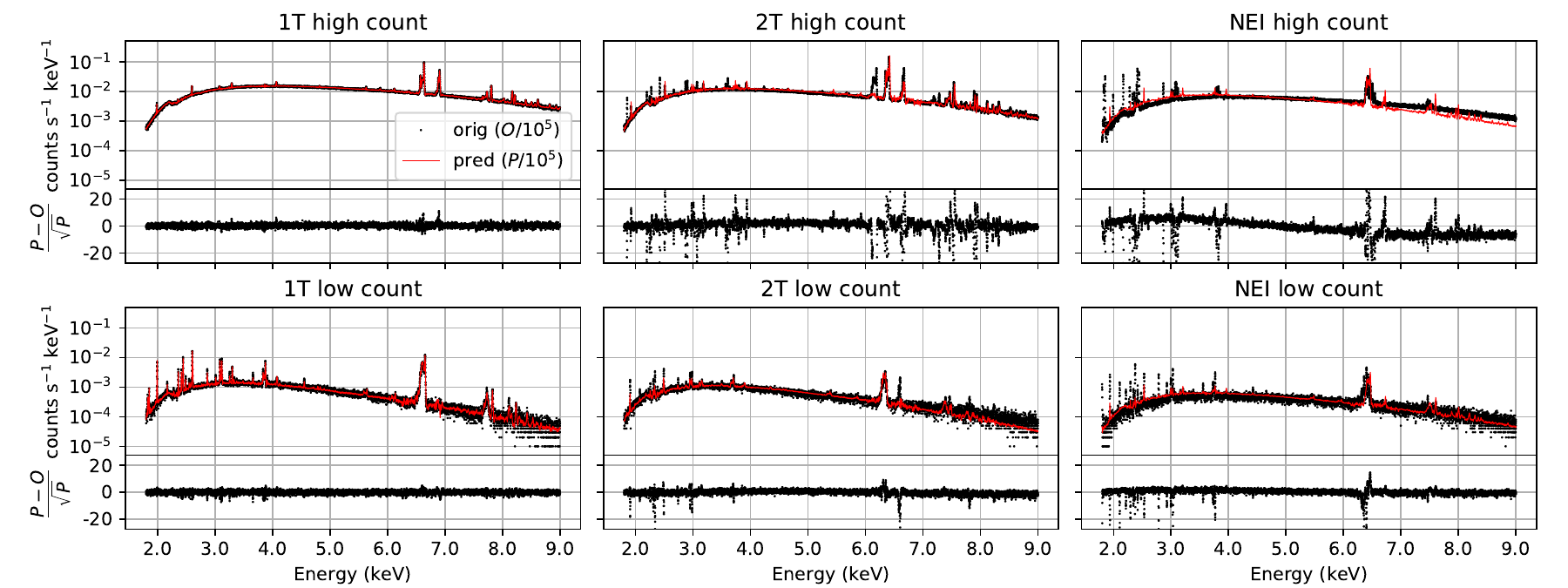}
 \caption[]{Examples of the VAE prediction for given observed spectra. The panels correspond to 1T plasma, 2T plasma and NEI plasma from left to right. The panels in the first and second rows correspond to high-count and low-count spectra, respectively. The normalized original data (black) and the corresponding predicted spectrum (red) are shown in each upper subpanel. The residual divided by the square-root of the prediction is shown in each lower subpanel.}
\label{img:comp_pred}
\end{figure*}
We feed the VAE network trained with the 1T training dataset with the 1T test dataset, 2T dataset, and NEI dataset. Figure~\ref{img:comp_pred} shows the examples of the VAE output for given data inputs. It is apparent that the output (red curve) is similar to the input spectrum (black points) in the cases of the 1T input (leftmost panels), whereas the input and output are considerably different in the 2T or NEI cases (middle and rightmost panels). Thus, by quantifying the difference between the input and output, one can automatically detect the input spectra which do not originate from the generative model of the training dataset. Note that the concept works both in the high-count (top panels) and low-count (bottom panels) cases. This illustrates the importance of using the appropriate loss function and activation function in the last layer (i.e. Poisson distribution and softplus function. See Section~\ref{sec:impl} and Eq.~\ref{eq:loss}).

There are several options regarding how to quantify the difference between the input and output. For demonstration, we simply choose the normalized chi-square statistic $\log\langle ((P-O)/\sqrt{P})^2\rangle$ for the metric, where $P$ (prediction) is the VAE output and $O$ (observation) is the VAE input.

\begin{figure}
  \centering
  \includegraphics[width=3.0in]{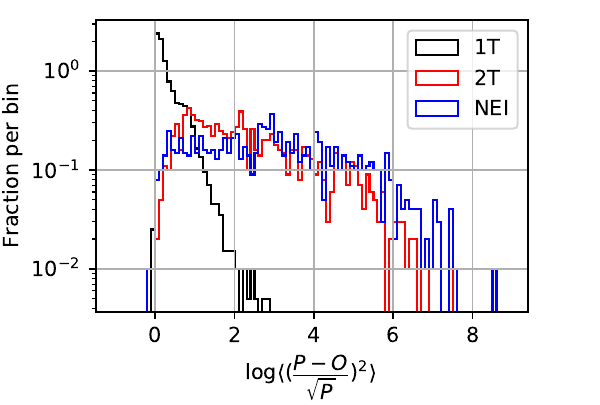}
 \caption[]{Distributions of $\log\langle ((P-O)/\sqrt{P})^2\rangle$ for 1T (black), 2T (red) and NEI (blue) datasets.}
\label{img:hist}
\end{figure}
Figure~\ref{img:hist} shows the distributions of the metric $M\equiv\log\langle ((P-O)/\sqrt{P})^2\rangle$ plotted for the 1T (black), 2T (red) and NEI (blue) datasets. While $M\lesssim 2$ for 1T dataset, $M>2$ for most of the 2T or NEI datasets. Therefore, for example, if the anomaly detection threshold is set to be $M=2$, most of the 2T and NEI data inputs are regarded as {\it anomalous}, while almost all of the 1T data inputs are regarded as {\it normal}.

\begin{figure}
  \centering
  \includegraphics[width=3.0in]{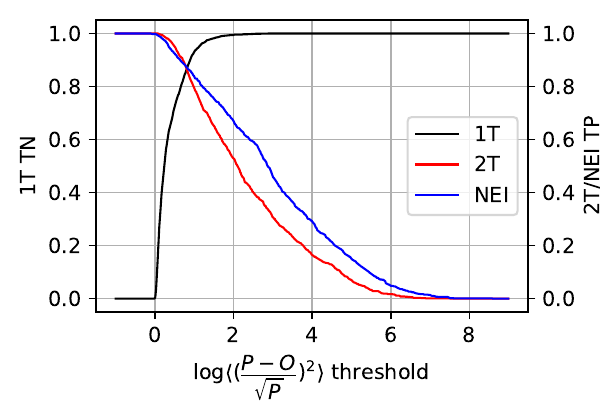}
 \caption[]{{\it Black:} the TN (true-negative) fraction for the 1T dataset, which is the fraction corresponding to the values below each given $M=\log\langle ((P-O)/\sqrt{P})^2\rangle$ value. {\it Red and blue:} the TP (true-positive) fraction for the 2T and NEI datasets, which is the fraction corresponding to the values above each given value of $M$.}
\label{img:performance}
\end{figure}
In general, lower anomaly detection thresholds result in more anomalies detected properly (true-positive; TP) and more samples falsely tagged as anomalies (false-positive; FP). On the other hand, higher thresholds gives more anomalies overlooked (false-negative; FN) and saves more normal samples (true-negative; TN). Figure~\ref{img:performance} illustrates this trade-off relation for our case. How these values change with respect to the threshold selection and the resulting optimal value of the threshold depend on the individual task and should be carefully determined.

\section{Discussion}\label{sec:discussion}
We have proposed and demonstrated a new anomaly detection technique for high-resolution X-ray spectroscopy. The method is based on the neural network architecture VAE, and only {\it normal} samples are used for the network training. Once the network is trained and the threshold is fixed, anomaly detection can be automatically performed.

Our method does not require anomalous samples for training. This is a significant advantage because one of the difficulties in using machine learning techniques (especially supervised ones) for anomaly detection is that it is often impossible to prepare the anomalous samples with a number sufficient for training. The difficulty originates mainly for the following two reasons; (1) anomalies by definition occur far less frequently than normal samples; (2) there are almost infinite varieties of anomalies, because normal samples occupy a small subspace of the entire input vector space and all others are anomalies.

The training of the VAE network takes time to a certain extent (several hours for the present case). On the other hand, network prediction using the trained network and subsequent computations are done almost instantly (less than a second for the present case). Therefore, by training the network in advance, it is easy to integrate the anomaly detection part into data processing pipelines. This might assist scientists in finding important things that would otherwise be missed due to the unmanageable amount of the data taken with future X-ray observatories.

One of the important aspects of VAE is that it simply approximates the generative model of the training samples, and the framework itself is independent of the nature of the data. Therefore, it is in principle possible to apply this method to other scenarios, e.g. detecting 3T plasma against 1T+2T samples or detecting non-thermal components out of thermal samples (e.g. in supernova remnants). However, the optimal training parameters such as the latent dimension of the network or the required number of simulated samples would be different from the current ones.

In Section~\ref{sec:results}, we used the anomaly metric of $M\equiv\log\langle ((P-O)/\sqrt{P})^2\rangle$ for demonstration purposes. The optimal metric will likely depend on what one wants to detect against what one regards as {\it normal} samples. For example, considering a very simple situation where the {\it anomaly} one wants to find is a single high-intensity line on a smooth continuum, the maximum of the residual (i.e. $\mathrm{max}(|P-O|)$) might outperform the normalized chi-square employed above. In general, different metrics yield different confusion matrix properties.

In the actual operation, in order to determine the optimal metric, one should also be able to plot a graph similar to Figures~\ref{img:hist} and \ref{img:performance} using a mock dataset simulated with the response files corresponding to the observation. Also, ROC AUC (the area under the receiver operating characteristic curve), is commonly used to select the optimal metric. That said, we think that the chi-square value would work reasonably well considering that statistical fluctuations will probably be significant almost always in high-resolution X-ray spectra.

The example metric value of $M=2$ is a very conservative choice for the current case. For example, if one wants to miss anomalies as little as possible, lower $M$ values might be preferred. On the other hand, if one wants to just point out extremely anomalous samples, higher $M$ values might work.

We investigated the properties of the samples that fall into the FN category, and found that some parameter conditions tend to result in lower values of $M$. For the 2T case, these conditions include for example (1) large normalization difference (i.e. $\mathrm{min}(N_1/N_2,N_2/N_1) < 0.15$), (2) low Fe abundance (i.e. $\mathrm{min}(Z_1,Z_2) < 0.3$), and (3) small redshift difference ($|z_1-z_2| < 0.015$), and we think all these conditions are intuitively understandable as the causes of the confusion. We also found that by excluding these `confusing' samples from the 2T dataset, TP ratio at $M=$1, 2, and 3 improved to $>0.9, \sim0.8$, and 0.6, respectively. On the other hand, we could not identify clear specific parameter conditions with which the NEI samples fall into the FN category. This might simply indicate the difficulty of distinguishing between NEI and equilibrium conditions possibly with the current response files.

In addition to our proposed methodology, we also tried a few different approaches to the anomaly detection also using the same VAE network. First, we compared the Mahalanobis distances in the latent space among the three cases as used in \citet{ghosh18}, but found no apparent difference between the single temperature case and the others. It is probably because the latent expressions are highly compressed and abstracted, and even if the information to distinguish anomalies is apparent in the raw spectrum, it is no longer evident in the latent space. Secondly, we observed the latent space using t-SNE \citep{maaten08}, which is another dimensional reduction technique for visualization of high-dimensional data. In the two-dimensional visualizations obtained using t-SNE, there seemed to be systematic differences among the data distributions of 1T, 2T, and NEI datasets. However, the visualized shape of each distribution is highly irregular and we found it difficult to quantify them.

We found that increasing the latent dimension does not make the t-SNE-visualized distributions clearer. In our setup, only four parameters are necessary to produce the input data (temperature, redshift, Fe abundance, and normalization). Thus, it is naturally expected that increasing the dimension of the latent space too much would not help the network gain higher expressive power. We examined the contribution rates in the latent space using principal component analysis, which resulted in four plus a few significant dimensions. This indicates that our choice of number of dimensions for the latent space, eight, was nearly optimal.

The anomaly detection also benefits the data preprocessing of X-ray spectroscopy, which we proposed in \citet{ichinohe18}. Assuming a certain spectral model, this method enables us to know the initial parameters for the spectral fitting at an accuracy of a few percent. While it works properly when the assumed spectral model matches the data, it would return meaningless values if the model does not describe the data well. From a practical point of view, when spectral fitting is attempted using models and initial parameters that are not suitable for the data, it will fail in various ways; the fitting may converge on a non-physical local minimum or may not converge at all with the optimizing process aborting. Such unwanted situations could be prevented beforehand using the anomaly detection method, with which one can separate out data that might require more careful treatment than usual. This makes the entire data processing procedure more robust.

In Section~\ref{sec:results}, we proposed an anomaly detection method in which the single anomaly detection metric $M$ is used. $M$ is computed using the information over the entire energy range. On the other hand, as illustrated using 2T and NEI datasets in Figure~\ref{img:comp_pred}, the VAE residuals of the anomalous samples can indicate individual spectral portions (or energy ranges) that the training model cannot reproduce. Therefore, by exploiting this more detailed information, it is even possible to point out the local anomalies in each spectrum.

Compared to the default one, this method is susceptible to local noise. This is shown in the upper leftmost panel in Figure~\ref{img:comp_pred} where some local residuals appear around 6.5--7~keV, even though this spectrum is a {\it normal} sample. However, despite the drawback, this might be useful for various applications. For example, it could be applied to the search of unknown line or absorption features, which has recently been gaining attention due to the possible sterile neutrino feature around 3.5~keV \citep{bulbul14,hitomidm}.

Although the current plasma codes are not perfect, they reasonably describe {\it Hitomi}'s spectra \citep[see e.g.][]{hitomiatomic}. One main reason is that all the calorimeter spectra of {\it Hitomi} were taken with the gate valve closed, which resulted in the spectral features below $\sim1.8$~keV (including the Fe--L complex) not being detected. The next X-ray mission {\it XRISM} is planned to be launched in the early 2020s and it is expected that the imperfectness of the plasma codes will come into focus as soon as {\it XRISM} starts observations with the gate valve open. Therefore, in the near future, the plasma codes will need to be modified so that they describe the data properly. Currently laboratory experiments to test plasma physics, such as X-ray spectroscopy in electron beam ion trap (EBIT) facilities, are ongoing \citep[e.g.][]{gall19}. Our proposed machine-learning anomaly detection would also help such activities by indicating the spectral parts which may need modifications.

Finally, it is intriguing to see how the current method works on the real spectra. When we apply the current method to the actual Hitomi spectrum \citep[Obs~3 all, see][]{ichinohe18}, we obtained $M=2.09$. While this at the first glance indicates the insufficiency of the 1T modelling of the cluster, we currently could however only say that the {\it Hitomi} Perseus spectrum of Obs~3 is not represented by a simple 1T model or it is not a point source.

Actually, there are also several other issues regarding applying the current method to the real spectral data. For example, while for demonstration we used only a single set of response files (RMF and ARF), in the actual situation, the response files may (especially ARFs) differ from pixel to pixel. Also, the contributions of e.g. CXB (cosmic X-ray background) and NXB (non-X-ray background) should be considered. Further works are certainly necessary in order to handle such realistic situations. We will cover in the future works the extensions of the present work which includes (1) applying the method to other scenarios in which {\it normal} and {\it anomalous} models are different from the current ones (i.e. 1T and 2T/NEI), and (2) dealing with the complexities arising in the actual situations.

\section*{Acknowledgements}
We thank the referee for constructive suggestions and comments. We thank Dr. Magnus Axelsson for helpful comments and English proofreading. YI is supported by Rikkyo University Special Fund for Research (SFR). This work was partially supported by MEXT-Supported Program for the Strategic Research Foundation at Private Universities, 2014-2019 (S1411024). This work was supported by JSPS KAKENHI Grant Numbers 15H05438. This work is supported by Astro-AI working group in RIKEN iTHEMS.









\bsp	
\label{lastpage}
\end{document}